\documentclass[%
 reprint, superscriptaddress, amsmath,amssymb, aps, pre
]{revtex4-1}

\usepackage{bm}% bold math
\usepackage{graphicx}% Include figure files
\usepackage{dcolumn}
\usepackage{xcolor}
\usepackage{comment}
\usepackage{amssymb}
\usepackage{amsmath}
\usepackage{esint}
\usepackage{epstopdf}
\usepackage{color}
\usepackage{gensymb}
\usepackage{mathtools}
\usepackage{appendix}
\usepackage{suffix}

\newcommand{\be}{\begin{equation}}
\newcommand{\ee}{\end{equation}}
\newcommand{\bea}{\begin{eqnarray}}
\newcommand{\eea}{\end{eqnarray}}

\newcommand{\br}{\mathbf{r}}

\newcommand{\e}{\varepsilon}

\newcommand{\bbr}{\bar{r}}
\newcommand{\bba}{\bar{a}}

\newcommand{\bn}{\bar{n}_{\rm cb}}

\newcommand{\bak}{\bar{k}}
\newcommand{\bap}{\bar{p}}

\newcommand{\p}{_{\rm p}}

\newcommand{\ce}{_{\rm c}}

\newcommand{\s}{_{\rm s}}
\newcommand{\w}{_{\rm w}}

\newcommand{\ew}{\varepsilon_{\rm w}}

\begin{document}
\preprint{APS/123-QED}

\title{Theoretical and computational analysis of the electrophoretic polymer mobility inversion induced by charge correlations}

\author{Xiang Yang}
\affiliation{Department of Applied Physics, Aalto University, P.O. Box 11000, FI-00076 Aalto, Finland}

\author{Sahin Buyukdagli}%
\affiliation{Department of Physics, Bilkent University, Ankara 06800, Turkey}%

\author{Alberto Scacchi}
\affiliation{Department of Applied Physics, Aalto University, P.O. Box 11000, FI-00076 Aalto, Finland.}
\affiliation{Academy of Finland Center of Excellence in Life-Inspired Hybrid Materials (LIBER), Aalto University, P.O. Box 16100, FI-00076 Aalto, Finland}

\author{Maria Sammalkorpi}
\affiliation{Academy of Finland Center of Excellence in Life-Inspired Hybrid Materials (LIBER), Aalto University, P.O. Box 16100, FI-00076 Aalto, Finland}
\affiliation{Department of Chemistry and Materials Science, Aalto University, P.O. Box 16100, FI-00076 Aalto, Finland}
\affiliation{Department of Bioproducts and Biosystems, Aalto University, P.O. Box 16100, FI-00076 Aalto, Finland}

\author{Tapio Ala-Nissila}
\affiliation{Department of Applied Physics, Aalto University, P.O. Box 11000, FI-00076 Aalto, Finland.}
\affiliation{Quantum Technology Finland Center of Excellence, Department of Applied Physics, Aalto University, P.O. Box 11000, FI-00076 Aalto, Finland}
\affiliation{Interdisciplinary Centre for Mathematical Modelling and Department of Mathematical Sciences, Loughborough University, Loughborough, Leicestershire LE11 3TU, UK}

\date{November 17, 2022}

\begin{abstract}
Electrophoretic (EP) mobility reversal is commonly observed for strongly charged macromolecules in multivalent salt solutions. This curious effect takes place, e.g., when a charged polymer, such as DNA, adsorbs excess counterions so that the counterion-dressed surface charge reverses its sign, leading to the inversion of the polymer drift driven by an external electric field. In order to characterize this seemingly counterintuitive phenomenon that cannot be captured by electrostatic mean-field theories, we adapt here a previously developed strong-coupling-dressed Poisson-Boltzmann approach to the cylindrical geometry of the polyelectrolyte-salt system. Within the framework of this formalism, we derive an analytical polymer mobility formula dressed by charge correlations. In qualitative agreement with polymer transport experiments, this mobility formula predicts that the increment of the monovalent salt, the decrease of the multivalent counterion valency, and the increase of the dielectric permittivity of the background solvent, suppress charge correlations and increase the multivalent bulk counterion concentration required for EP mobility reversal. These results are corroborated by coarse-grained molecular dynamics simulations showing how multivalent counterions induce mobility inversion at dilute concentrations and suppress the inversion effect at large concentrations. This re-entrant behavior, previously observed in the aggregation of like-charged polymer solutions, calls for verification by polymer transport experiments.
\end{abstract}

%\keywords{Suggested keywords}
\maketitle

\section{Introduction}
Electrostatic correlation effects are ubiquitous in biological systems involving strongly charged biomolecules and membranes.
Counterintuitive phenomena, such as like-charge attraction, are typically observed in systems including macromolecules in contact with multivalent counterions~\cite{Grosberg2002,Levin2002,Angelini2003,Besteman2007,Besteman2004,Butler2003}.
Several mechanisms mediating the attraction between like-charged rods have been considered in literature, such as, covalence-like binding~\cite{Ray1997}, Gaussian-fluctuation correlations~\cite{Podgornik1998}, and structural correlations~\cite{Kornyshev1997,Kornyshev1999}. 
In dense polymer systems, like-charge attraction can induce, e.g., bundle formation of F-actin and toroidal aggregates of DNA ~\cite{Tang1996,Bloomfield1997}. Multivalent inorganic ions and polyamines can also act as condensing agents~\cite{Teif2011}. 

Charge inversion (CI) is another interesting manifestation of charge correlations. This phenomenon occurs when the macromolecular surface charge flips its sign upon the adsorption of a sufficient amount of counterions from the solution. When combined with other physical factors, CI can induce additional effects. For example, in a press-driven flow through a negatively charged slit pore, multivalent cation addition can invert the sign of the monovalent counterion current, generating a like-charged streaming current of negative sign~\cite{VanderHeyden2006}. 
One should also mention the important role played by CI in biology. In the cell medium of eukaryotic organisms, the negatively charged DNA and the positively charged histone proteins can assemble into nucleosome. At high salt concentrations, the stable structure corresponds to a rope-like DNA wrapped around a bead-like histone having an inverted net charge~\cite{Rippe2008,Grosberg2002}. 

Electrophoresis is the motion of dispersed charged particles relative to a fluid exposed to an electric field.
The resulting driven transport is quantified in terms of the electrophoretic (EP) mobility, defined as the ratio of the drift velocity and the electric field strength. Based on the EP mobility strength, one can separate macromolecules in terms of their size and surface charge~\cite{Kaper2003,danger2007}. 
Furthermore, EP transport experiments can be efficiently used to probe the interfacial charge structure of polyelectrolytes~\cite{Kabanov1995,Kabanov1996}. %,such as the mentioned CI 
Additionally, from viral infection to nanopore-based polymer sensing, polymer transport through confined pores plays a critical role in biological processes and various nanoscale applications~\cite{Holm2001,Schoch2008,Jain2016,Thomas2016}. The underlying electrohydrodynamic mechanisms have been extensively explored in recent theoretical works~\cite{Buyukdagli2020,Buyukdagli2021}. 

The surface CI of polyelectrolytes has been previously characterized by EP transport experiments~\cite{Wang2016,Wang2018}. Indeed, these experiments have shown that the addition of multivalent cations into a polyelectrolyte solution induces the EP motion of the negatively charged polyelectrolytes along the external electric field. In this article, we carry out a theoretical investigation of this seemingly counterintuitive mobility reversal effect. As the occurrence of this phenomenon requires the presence of multivalent counterions strongly coupled to the polyelectrolyte charges, the characterization of the underlying mechanism necessitates the use of a correlation-corrected electrostatic framework. Thus, considering the linear response limit of a previously developed strong coupling (SC)-dressed Poisson-Boltzmann (PB) approach~\cite{Kanduc2010,Kanduc2011,Buyukdagli2020,Buyukdagli2021}, we derive an analytical polymer transport formula accounting for the multivalent ion-induced charge correlations. 

We show that our correlation-corrected EP transport formula can qualitatively reproduce various experimentally observed correlation effects on polymer transport, such as the onset of the mobility reversal by added multivalent counterions, and its weakening by the increment of the monovalent salt component, the increase of the solvent permittivity, and the reduction of the multivalent counterion valency. The analytical structure of our formalism enables a clear interpretation of the correlation mechanism driving these effects~\cite{Wang2016,Wang2018}. 

As a complementary approach, we also perform particle-based coarse-grained molecular dynamics (MD) simulations. A benchmark between MD simulations and the dressed-ion theory is crucial, since the latter has been previously tested for solutions in contact with charged planes~\cite{Kanduc2010,Kanduc2011,Buyukdagli2020,Buyukdagli2021}, but not in the case of charged cylinders. 
In addition to reproducing consistently the aforementioned theoretical predictions, in the dense multivalent counterion regime, our MD simulations capture a re-entrant phase not covered by the theoretical mobility formula, which is valid only at dilute counterion densities. Finally, within our computational framework, we also characterize the effect of the counterion charge distribution beyond the point-ion approximation. The limitations of our theory and potential improvements are discussed in the {\it Summary and Conclusions} section.

\section{Molecular Dynamics simulation approach} 
\label{MD}
\subsection{Simulation system}

We start by discussing the details of the MD simulations in this work. The simulations were run in a box of volume $L_x\times L_y\times L_z$, where the longitudinal dimension was set to $L_z= 20$~nm. Depending on the added salt concentration, the transverse dimensions $L_x=L_y$ were altered between 24~nm and 240~nm. As explained in the Supplementary Material (SM), we chose $L_z$ to obviate finite-size effects from the data.

\begin{figure}[th]
\begin{centering}
\includegraphics[width=0.95\linewidth]{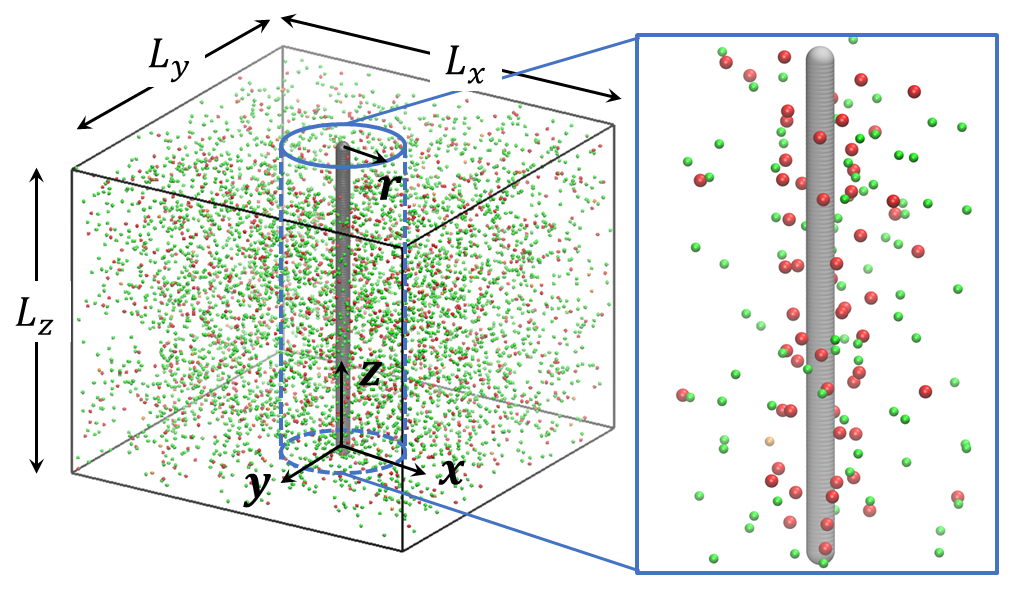} 
\par\end{centering}
\caption{Left panel: A schematic of the CG DNA located at $(x,y)=(0,0)$ along the $z$ axis of the simulation box with total volume $L_x\times L_y\times L_z$. Right panel: Ions distributed within a cylindrical volume of radius $r$ centered at $(x,y)=(0,0)$. The red, green and orange spheres represent trivalent cations, monovalent anions and monovalent cations, respectively. In this snapshot, $L_x=L_y=24$~nm, and $L_z=20$~nm. The concentration of multivalent salt (TCl$_3$) is $n_{\rm cb}=200$~mM and that of monovalent counterions (e.g. Na$^+$) neutralizing the CG DNA charge 17~mM.
\label{fig:system}}
\end{figure}

As depicted in Fig.~\ref{fig:system}, the system includes a coarse-grained (CG) DNA chain centered at the coordinates $(x,y)=(0,0)$ spanning the cuboid simulation box along the $z$ axis. The DNA molecule was modeled by a series of spherical beads distributed uniformly along the polymer axis and separated by $b=0.17$~nm from each other. The value of $b$ is small enough to provide a smooth potential surface in the longitudinal direction. Moreover, each bead carries an elementary charge $-e$, resulting in a linear charge density of $\lambda_0= -e/b = -5.9\;e$/nm. The radius of the polymer, or equivalently the radius of the beads, was set to $a=1.2$~nm, which corresponds to the characteristic thickness of double-stranded DNA molecules~\cite{Forrey2006,Mandelkern1981}.

Here the simulations were in the canonical ensemble, i.e. with fixed number of particles. Therefore, electroneutrality in our simulations was achieved by neutralizing the DNA charges with  $L_z/b$ added monovalent counterions (e.g. Na$^+$). This lead to a counterion concentration of $(L_x L_y b)^{-1}$. % amount
Additional monovalent salt, such as NaCl, and multivalent salt species of general chemical structure XCl$_{q_{\rm c}}$, were added into the solution. The different cations (X) of equal size and valency $q\ce$ were divalent, trivalent (T$^{3+}$), quadrivalent or octavalent charges.
Within the theory considerations, the bulk concentration of the added monovalent and multivalent ions will be denoted by $n_{\rm sb}$ and $n_{\rm cb}$, respectively.

Overlapping of the mobile charges was avoided by placing the ions on a regular grid at the initialization stage. The pairwise interactions between the polymer beads and the ions were modelled via the standard Weeks-Chandlers-Andersen potential~\cite{Weeks1971}
\begin{equation}
       V_{ij}(r)= \left\{4 \epsilon_{ij} \left[\left(\frac{\sigma_{ij}}{r}\right)^{12} -\left(\frac{\sigma_{ij}}{r}\right)^6 \right]+\epsilon_{ij}\right\}\theta(r_{ij}^{\rm cut}-r),
\label{eq:WCA1}
\end{equation}
where the indices $i$ and $j$ correspond to the multivalent, sodium, chloride ions, and DNA beads; $\theta(x)$ is the Heaviside step function. In the implementation of the potential~(\ref{eq:WCA1}), we used the Lorentz-Berthelot mixing rules $\sigma_{ij}=(\sigma_{i}+\sigma_{j})/2$ and $\epsilon_{ij}=\sqrt{\epsilon_{i}\epsilon_{j}}$,  and set the cut-off radii of the pairwise interactions $r_{ij}^{\rm cut}=2^{1/6}\sigma_{ij}$. The parameters $\epsilon_{i}$ and $\sigma_{i}$ are reported in Table~\ref{tab:WCA}. The approach follows our previous work, see Ref.~\cite{Vahid2022} for more details.

The pairwise electrostatic interactions between charges $q_i$ and $q_j$, with respective position vectors $\br_i$ and $\br_j$, were taken into account with the Coulomb potential
\begin{equation}
V^{\rm c}(\br_i-\br_j)=\frac{q_iq_je^2}{4\pi \varepsilon_{\rm w}||\br_i-\br_j||}.
\label{co}
\end{equation}
These potentials treat the solution as a continuous dielectric medium with permittivity $\ew$. Unless stated otherwise, the relative dielectric constant of the Coulomb potential was set to the relative permittivity of water $\ew/ \varepsilon_0=78$, where $\varepsilon_0$ stands for the vacuum permittivity.
The MD simulations were performed using the LAMMPS Jan2020 package~\cite{Plimpton1995,Thompson2022}.

The long-range electrostatic interactions were calculated with the particle-particle particle-mesh (PPPM) method~\cite{Hockney2021}. Up to the characteristic split distance $d^{\rm s}$, the pairwise Coulomb interactions in Eq.~(\ref{co}) were evaluated in real space, whereas beyond this distance in reciprocal space. The PPPM accuracy parameter (relative error) was set to $10^{-5}$ and the stencil size parameter to 5. The setting of the distance $d^{\rm s}$ is discussed in the SM.
All simulations were performed in the $NVT$ ensemble. During the simulations, the temperature was controlled by the Nose-Hoover thermostat with $300$ K as the reference temperature~\cite{Nose1984,Hoover1985}. After an initial placement of the ions into the simulation box, the system energy was minimized by the conjugate gradient method. This was followed by a 24 ns $NVT$ simulation run for data analysis. The first 4 ns of this run were disregarded. A 2 fs time step was used. 

\begin{table}[ht]
\begin{centering}
\begin{tabular}{||c |c |c |c | c||}
\hline
Variable & DNA bead & Na$^+$ & Cl$^-$ & multivalent ion \tabularnewline
\hline
\hline
$\epsilon$ (kcal/mol) & 0.1 & 0.13 & 0.124 & 0.1\tabularnewline
\hline
$\sigma$ (nm) & 2.4 & 0.234 & 0.378 & 0.5\tabularnewline
\hline
\end{tabular}
\par\end{centering}
\caption{The parameters of the Weeks-Chandlers-Andersen potential used in the CG model.
\label{tab:WCA}}
\end{table}

\subsection{DNA Charge Inversion}

In our DNA-liquid system, the mobile charge configuration is set by the collective effect of the electrostatic and steric interactions between the ions and the DNA beads, and thermal fluctuations suppress the electrostatic coupling of the charged entities. It has been previously shown that in the presence of a sufficient amount of multivalent counterions in the liquid, the electrostatic many-body effects take over the thermal fluctuations, leading to the overcompensation of the bare DNA charges by the counterions bound to the DNA molecule~\cite{Netz2001,Wang2016,Wang2018,Hsiao2008,Kanduc2011,Buyukdagli2021,Aksimentiev2010}. 

\begin{figure}
\begin{centering}
\includegraphics[width=0.95\linewidth]{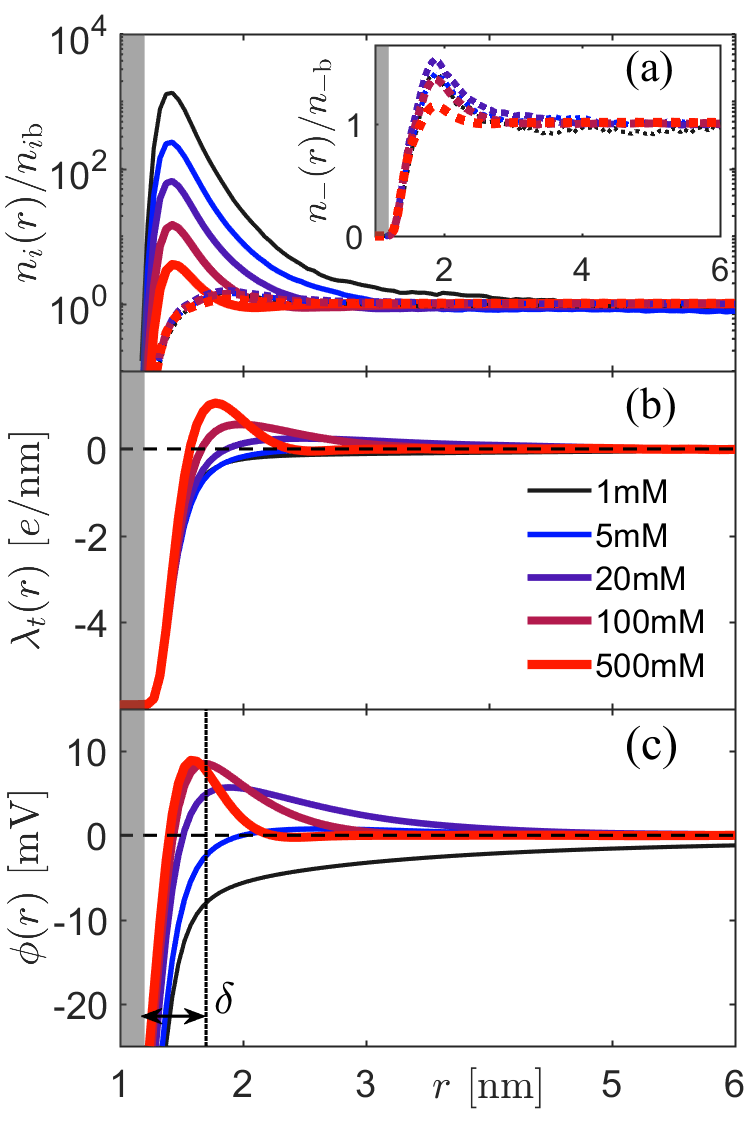}
\par\end{centering}
\caption{Radial functions related to charge inversion from the MD simulations. (a) Normalized density profiles of the T$^{3+}$ cations (solid curves) and Cl$^{-}$ anions (dashed curves). (b) Cumulative charge density in Eq.~(\ref{eq:cumChar}). (c) Electrostatic potential in Eq.~(\ref{eq:potential}) at various trivalent salt concentrations $n_{\rm cb}$ indicated in the legend of (b). The inset in (a) displays the dimensionless Cl$^{-}$ densities on a smaller linear scale. The dashed vertical line in (c) marks the no-slip boundary located at $r=a+\delta$. The gray regions indicate the radius of the polymer $a$. The relative liquid permittivity is $\ew/ \varepsilon_0=78$.
\label{fig:dens}}
\end{figure}

With the aim of illustrating the corresponding CI effect, and of relating the latter to the EP mobility inversion of the molecule, we focus first on the configuration of trivalent salt (TCl$_3$). %at the water permittivity $\varepsilon_{\rm w}/ \varepsilon_0=78$.
Figure~\ref{fig:dens}(a) displays the dimensionless density profiles $n_{i}(r)/n_{i\rm{b}}$ of the multivalent cations (solid curves) and monovalent anions (dashed curves) at various bulk TCl$_3$ concentrations from the MD simulations. The density profiles $n_{i}(r)$ are calculated from counting the ions in a cylindrical volume around the DNA (cf. Fig.~\ref{fig:system}). One can see that the T$^{3+}$ adsorption peaks, located at $r\approx 1.4$ nm, are followed by the Cl$^{-}$ density peaks at $r\approx 1.8$ nm. Moreover, the inset shows that these peaks correspond to the interfacial excess of Cl$^{-}$ ions. Thus, the Cl$^{-}$ attraction by the adsorbed T$^{3+}$ cations leads to an apparent like-charge binding of the Cl$^{-}$ anions onto the anionic DNA molecule.

The DNA zeta potential setting the EP mobility of the polymer can be obtained from the cumulative charge density $\lambda_t(r)$ of the molecule. The latter corresponds to the total charge enclosed by a cylindrical volume of radius $r$ divided by the length of the cylinder, $L_z$ (see Fig.~\ref{fig:system}), i.e.
\begin{equation}
    \lambda_t(r)= \lambda_0 + 2\pi e\int_{0}^{r} dr'r'[n_{\rm +}(r')-n_{\rm -}(r')+q_{\rm c}n_{\rm c}(r')] ,
\label{eq:cumChar}
\end{equation}
where $n_{+}(r)$, $n_{-}(r)$ and $n_{c}(r)$ denote the number density of the monovalent Na$^{+}$ and Cl$^{-}$ ions, and the multivalent X$^{q\ce+}$ cations, respectively. Figure~\ref{fig:dens}(b), displaying the cumulative charge density in Eq.~(\ref{eq:cumChar}), shows that CI occurs in the bulk T$^{3+}$ concentration regime $n_{\rm cb}\gtrsim20$ mM.

\begin{figure*}[th]
\begin{centering}
\includegraphics[width=1.0\linewidth]{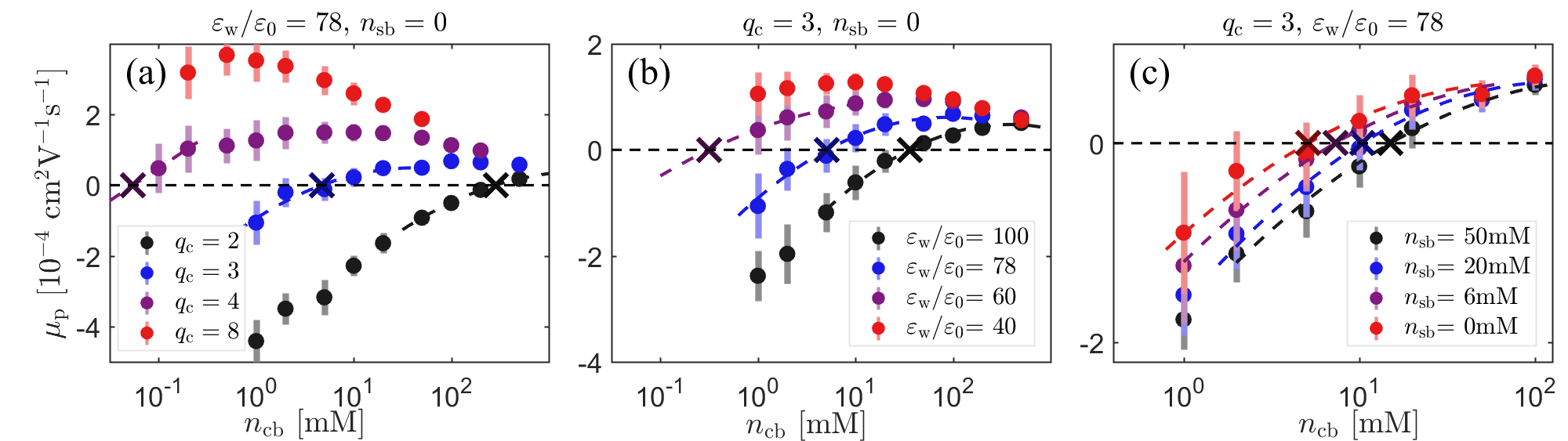}
\par\end{centering}
\caption{EP mobility $\mu\p$ from MD simulations as a function of the multivalent counterion concentration $n_{\rm cb}$ for various values of (a) the cation valency $q\ce$, (b) the dielectric constant, and (c) the monovalent salt concentration $n_{\rm sb}$. The dashed curves correspond to a quadratic fitting function (see main text), and the crosses represent the reversal concentrations $n_{\rm cb}^*$, which are reported in Table~\ref{tab:n_cb}.
\label{fig:mobiCurve}}
\end{figure*}

By integrating the radial Poisson equation, one can relate the radial component of the electric field $E(r)$ to the cumulative charge density as $E(r)= \lambda_t(r)/2\pi r\ew$.
Upon integration of the latter equality, the electrostatic potential $\phi(r)$ takes the form
\begin{equation}
\phi(r)= \int_{r}^{\infty} \frac{\lambda_t(r')}{2\pi\ew r'}dr'.
\label{eq:potential}
\end{equation}
The resulting potential profiles displayed in Fig.~\ref{fig:dens}(c) show that CI manifests itself as the emergence of the electrostatic potential peaks, where the slope of $\phi$ switches from positive to negative.

\begin{comment}
\begin{figure}[th]
\begin{centering}
\includegraphics[width=0.95\linewidth]{Figs/mobiCurves_lowXi.png}
\par\end{centering}
\begin{centering}
\includegraphics[width=0.95\linewidth]{Figs/mobiRev_lowXi.png}
\par\end{centering}
\caption{Dimensionless reversal concentration $\bar{n}_{\rm cb}^*$ as a function of the coupling parameter $\Xi$. The colors of data stand for screen parameter $s$.
{\color{blue}combine this figure to Fig. 3 and 4...}
\label{fig:mobiRev_lowXi}}
\end{figure}
\end{comment}

As described in Sec.~\ref{th}, the coupled solutions of the electrostatic Poisson equation and the hydrodynamic Stokes relation yield the Helmholtz-Smoluchowski identity, which relates the EP polymer mobility to the DNA zeta potential $\zeta$ via
\begin{equation}
    \mu\p=\frac{\ew \zeta}{\eta},
\label{eq:Smol}
\end{equation}
where $\eta=8.91\times 10^{-4}$ Pa s is the dynamic viscosity of water. In Eq.~(\ref{eq:Smol}), the zeta potential corresponds to the electrostatic potential value at the no-slip surface separating the mobile ions from the ones bound to the charged polymer~\cite{Bhattacharjee2016}. Given this definition, the zeta potential can be obtained from our potential profiles as $\zeta= \phi (a+\delta)$. 
Based on the results of previous experiments~\cite{Yamaguchi2016,Galla2014}, in our calculations the thickness of the no-slip region was set to $\delta=0.5$ nm.

\subsection{Electrophoretic mobility inversion}

We next focus on the EP mobility under a variety of conditions from the MD simulations. %All of them are calculated from ion distribution with Eq. (\ref{eq:cumChar})-(\ref{eq:Smol}).
Figure~\ref{fig:mobiCurve}(a) illustrates the dependence of the EP mobility on the multivalent cation at concentration $n_{\rm cb}$ for various valencies. First, the plot shows that for $q\ce=2-4$ the increment of the concentration $n_{\rm cb}$ increases the negative polymer mobility and switches it to positive. The characteristic concentration $n_{\rm cb}^*$ at which the mobility is inverted can be identified by a quadratic fit of the curve $\mu\p(n_{\rm cb})=0$ in the vicinity of the inversion point. In Fig.~\ref{fig:mobiCurve} these fitting functions are shown by dashed lines. The reversal concentrations provided by the fits are $n_{\rm cb}^*=280$, $5$, and $0.05$ mM, for divalent, trivalent, and quadrivalent salt, respectively. This is in agreement with prior findings on asymmetricity of the salt and charge reversal~\cite{Antila2017}. Additionally, in agreement with polymer transport experiments~\cite{Wang2016,Wang2018}, the increase of the counterion valency lowers the critical salt concentration $n_{\rm cb}^*$ required for the occurrence of the mobility reversal, i.e. $q\ce\uparrow n_{\rm cb}^*\downarrow$. 

The intensification of the DNA CI and the resulting mobility reversal by ion valency stems from the amplification of charge correlations. Indeed, in Sec.~\ref{sc}, we show that the weight of the charge correlations responsible for CI is proportional to the electrostatic coupling parameter defined as $\Xi=q\ce^2\ell_{\rm B}/\mu_{\rm GC}$, where $\ell_{\rm B}=e^2/(4\pi\ew k_{\rm B}T)$ stands for the Bjerrum length, and $\mu_{\rm GC}=1/(2\pi q\ce\ell_{\rm B}\sigma\p)$ is the Gouy-Chapman (GC) length. Here
$\sigma\p$ is the surface-charge number density of the polymer, and $\sigma\p=1/2\pi ab$ in the simulations. These identities imply that the coupling parameter is a cubic function of the ion valency ($\Xi\propto q\ce^3$), explaining the sharp emergence of mobility reversal upon the rise of the counterion valency at constant concentration $n_{\rm cb}$, as can be seen in Fig.~\ref{fig:mobiCurve}(a). 

Figure~\ref{fig:mobiCurve}(a) also shows that for $q\ce\geq2$, the EP mobility exhibits a non-monotonic dependence on $n_{\rm cb}$. Namely, as the multivalent cation concentration is increased beyond the value $n_{\rm cb}^*$, the reversed mobility rises, reaches a peak, and drops monotonically. This peculiarity is qualitatively similar to the {\it re-entrance} phenomenon observed in like-charge polymer interactions; according to prior experimental studies and theoretical analysis of these systems, multivalent cations triggering DNA condensation at low concentrations reverse the effect at large concentrations, resulting in the segregation of the condensates in the charged liquid~\cite{Hsiao2008,Buyukdagli2017,Truzzolillo2018}.

The permittivity of the solvent is an additional control parameter previously investigated by transport experiments~\cite{Wang2016}. The effect of the liquid permittivity on the EP polymer velocity is displayed in Fig.~\ref{fig:mobiCurve}(b). One can see that at fixed multivalent ion concentration, the reduction of the solvent permittivity rises the negative DNA mobility and switches the latter to positive, i.e. $\ew\downarrow\mu\p\uparrow$. This implies that the mixing of the water solvent with a lower permittivity liquid can solely trigger DNA mobility inversion. This result is in agreement with polymer transport experiments in Ref.~\citenum{Wang2016}, where the negative DNA velocity in water-ethanol mixtures was observed to rise and reverse with increasing the volume fraction of ethanol in the liquid.

The occurrence of mobility reversal upon the reduction of the solvent permittivity is a consequence of stronger charge correlations in lower permittivity liquids. Indeed, according to its definition above, the electrostatic coupling parameter scales quadratically with the inverse dielectric permittivity, i.e. $\Xi\propto\ew^{-2}$. Hence, the reduction of the liquid permittivity enhances the weight of the electrostatic many-body interactions responsible for CI and mobility reversal. Owing to this mechanism, Fig.~\ref{fig:mobiCurve}(b) shows that the multivalent ion concentration at the reversal drops with the liquid permittivity, i.e. $\ew\downarrow n_{\rm cb}^*\downarrow$.

Finally, in Fig.~\ref{fig:mobiCurve}(c), we investigate the effect of added monovalent salt, such as NaCl, at concentration $n_{\rm sb}$, on DNA mobility. The plot indicates that monovalent salt ions counteract the multivalent cations and monotonically suppress the mobility inversion. Consequently, in accordance with polymer transport experiments~\cite{Wang2016,Wang2018}, the minimum counterion concentration for mobility reversal increases with the amount of monovalent salt, i.e. $n_{\rm sb}\uparrow n^*_{\rm cb}\uparrow$. In Sec.~\ref{th}, within the framework of our correlation-augmented EP transport theory, we show that this feature originates from the attenuation of charge correlations by monovalent ions.

\begin{table}[ht]
\begin{centering}
\begin{tabular}{||p{3.2cm} |p{0.8cm} |p{0.8cm} |p{0.8cm} |p{0.8cm}||}
\hline
Parameters & \multicolumn{4}{c||}{$n^*_{\rm cb}$ [mM]} \tabularnewline
\hline
Fig. 3 (a) & & & & \tabularnewline
$q\ce=2,3,4,8$ & n/a & 0.055 & 5 & 280 \tabularnewline
\hline
Fig. 3 (b) & & & & \tabularnewline
$\ew/ \varepsilon_0=100,78,60,40$ & n/a & 0.3 & 5 & 35\tabularnewline
\hline
Fig. 3 (c) & & & & \tabularnewline
$n_{\rm sb}=0,6,20,50$~mM & 5 & 7.2 & 10 & 15\tabularnewline
\hline
Fig. 4 & & & & \tabularnewline
rod A, B, C, D & 70 & 80 & 140 & 260 \tabularnewline
\hline
\end{tabular}
\par\end{centering}
\caption{The reversal concentrations $n^*_{\rm cb}$ from MD simulations for different parameters in Figs. 3 and 4. n/a indicates that $n^*_{\rm cb}$ is too small to be reliably estimated from the simulations.
\label{tab:n_cb}}
\end{table}

\subsection{Influence of charge distribution in multivalent counterions}

In order to extend our understanding of the mechanism behind mobility inversion beyond the counterion valency, we scrutinize the role played by the spatial charge distribution in the multivalent counterions on polymer mobility. To this end, we ran MD simulations by replacing the spherical trivalent counterions by rod-like charges with the same valency. Each rod-like ion of total length $L=1.6$~nm consists of seven spheres linearly distributed. The diameter of each sphere is $0.5$ nm, same as previous multivalent spherical ions (see Table I), and the distance between two adjacent spheres is $0.18$~nm. 

\begin{figure}[th]
\begin{centering}
\includegraphics[width=0.95\linewidth]{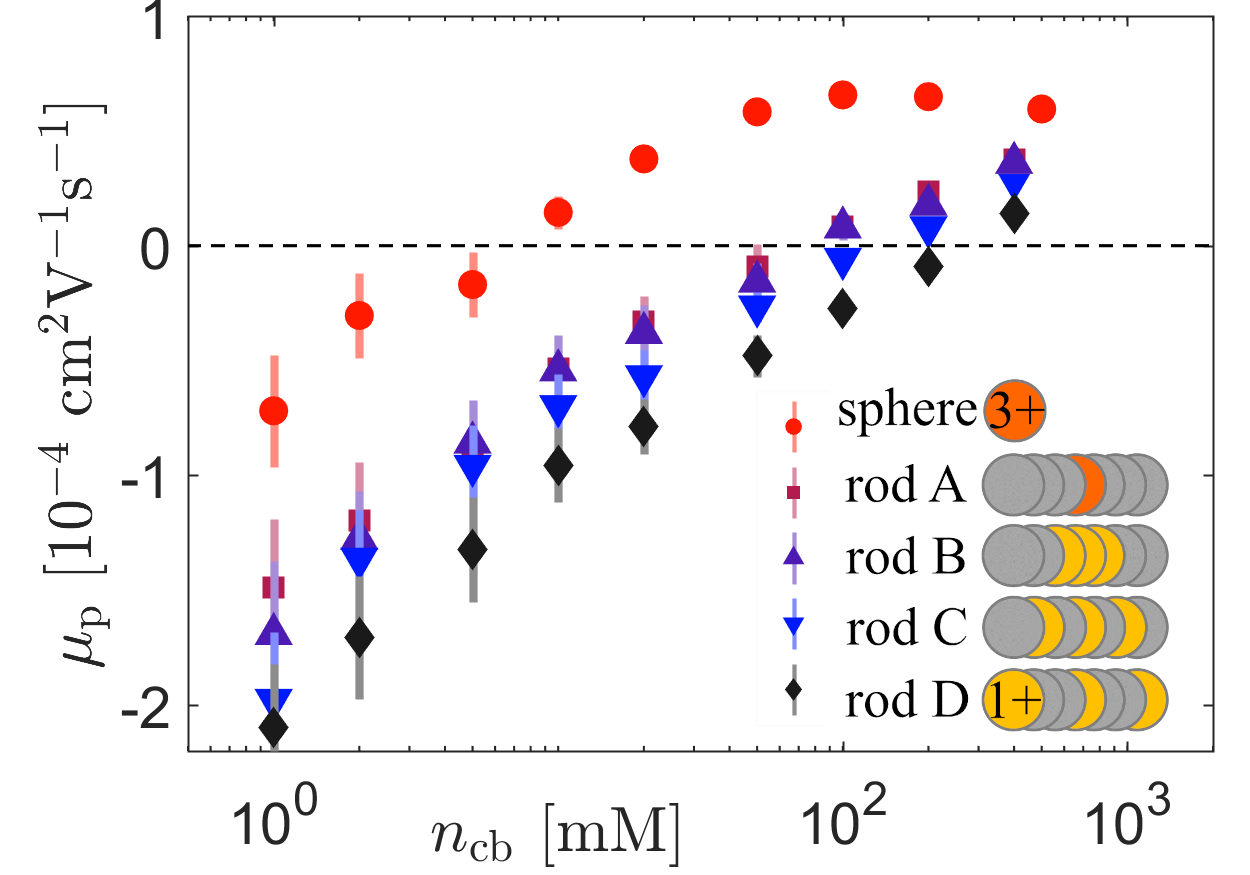}
\par\end{centering}
\caption{EP mobility from MD simulations in the presence of trivalent counterions of different intramolecular structures.
The reversal concentrations are found at $n_{\rm cb}^*=7, 70, 80, 140$ and 260~mM for sphere, rod A, B, C and D, respectively. The reversal concentrations $n_{\rm cb}^*$ are shown in Table~\ref{tab:n_cb}. Here the concentration of monovalent salt is $n_{\rm sb}=0$. In the inset the red circle denotes a spherical trivalent counterion, whereas the yellow circles are monovalent spherical counterions located at different sites in the seven-bead rodlike molecule.
\label{fig:mobiConf}}
\end{figure}

First, we analyze the purely steric effect originating from the finite length of the trivalent ions. To this aim, we place three elementary charges on the central sphere of the rod (rod A in Fig.~\ref{fig:mobiConf}), and compare the resulting polymer mobility (cyan curve) to the case with spherical trivalent ions (red curve). The plot indicates that as the rotational penalty experienced by the rod-like ions close to the polymer surface reduces their density and their degree of condensation, the finite counterion size decreases the polymer mobility whilst increasing the critical counterion concentration required for mobility inversion from $n_{\rm cb}^*=7$ mM to $n_{\rm cb}^*=70$ mM.

Second, we investigate the electrostatic effect associated with the surface charge density of the multivalent counterions. To this purpose, we split the three elementary charges on the ion by moving two unit charges from the center to the end in a symmetric fashion (see the corresponding configurations in the legend of Fig.~\ref{fig:mobiConf}). One can see that the resulting reduction of the ionic surface charge density weakens the DNA screening by the counterions, lowering the EP polymer mobility monotonically at all concentrations. As a result, the critical ion concentration for mobility reversal rises from $n_{\rm cb}^*=70$ mM for rod A-like counterions to $n_{\rm cb}^*=80$ mM, $140$ mM, and $260$ mM for rod-like counterions of type B, C, and D, respectively.

\section{SC-dressed EP transport theory}
\label{th}

In order to physically understand the various aspects of the EP mobility reversal investigated in the previous section by MD simulations, we develop here a SC-corrected analytical theory of EP polymer transport. The theory is based on the dressed-ion limit~\cite{Kanduc2011} of the SCPB formalism~\cite{Buyukdagli2020,Buyukdagli2021}, treating the monovalent salt ions at the weak-coupling level while taking into account the SC correlations mediated by the multivalent charge species. We emphasize that due to its grand-canonical nature, the present theory can consistently account for the bulk charge reservoir present in the real system. This feature leads into quantitative differences between the theory and the canonical MD simulations here. In contrast to the MD simulations, where the finite number of charges in the setup required to force the electroneutrality condition via the addition of extra monovalent cations, in our grand-canonical formalism the inclusion of the chemical equilibrium between the interfacial and bulk charges allows to satisfy automatically both the DNA and the bulk electroneutrality conditions.

\subsection{Derivation of the SC-dressed polymer mobility}

The EP mobility of a cylindrical polymer with radius $a$ translocating through a nanopore of radius $d$ has been previously calculated by the coupled solution of the Navier-Stokes and Poisson equations in Ref.~\cite{Buyukdagli2021}. Taking the limit $d\to\infty$ to remove the membrane interface from the model, the EP polymer mobility in the bulk liquid follows from this previous result as
\be
\label{eq0}
\mu\p=\mu_{\rm ep}\left[\phi(a^*)-\phi(d\to\infty)\right],
\ee
where $\phi(\br)=eV(\br)/(k_{\rm B}T)$ is the dimensionless electrostatic potential. Moreover, we introduced the EP mobility coefficient defined as $\mu_{\rm ep}=\ew\e_0k_{\rm B}T/(e\eta)$, with the dielectric permittivity of vacuum $\e_0$ and water $\ew$, the thermal energy $k_{\rm B}T$, the electron charge $e$, and the water viscosity $\eta$. In Eq.~(\ref{eq0}), the effective polymer radius $a^*$ is related to the physical polymer radius $a$ by $a^*=a+\delta$, where $\delta$ is the hydrodynamic no-slip length. We also note that the relation~(\ref{eq0}) is naturally equivalent to Eq.~(\ref{eq:Smol}).

The evaluation of the mobility~(\ref{eq0}) requires the calculation of the average potential. In the dressed-ion limit~\cite{Kanduc2011} of the SCPB formalism~\cite{Buyukdagli2020} describing the electrostatics of multivalent electrolyte mixtures, the SC-dressed average electrostatic potential is given by
\be\label{eq1}
\phi(\br)=\phi\s(\br)+\phi\ce(\br),
\ee
where the average potential component associated with the monovalent salt and satisfying the linear PB equation reads
\be\label{eq2}
\phi\s(\br)=\int\mathrm{d}^3\br' G(\br,\br')\sigma(\br'),
\ee
and the SC potential induced by the multivalent-ion component of valency $q\ce$ and bulk concentration $n_{\rm cb}$ is
\be\label{eq3}
\phi\ce(\br)=q\ce n_{\rm cb}\int\mathrm{d}^3\br' G(\br,\br')k\ce(\br').
\ee
In Eqs.~(\ref{eq2})-(\ref{eq3}), we introduced the fixed surface charge of the polymer $\sigma(\br)=-\sigma\p\delta(r-a)$, and the electrostatic Green function satisfying the kernel equation
\be\label{eq4}
\left[\nabla\cdot\e(\br)\nabla-\kappa^2\e(\br)\theta(r-a)\right]G(\br,\br')=-\frac{e^2}{k_{\rm B}T}\delta(\br-\br').
\ee
In Eq.~(\ref{eq4}), we defined the radial dielectric permittivity profile
\be
\label{eq5}
\e(\br)=\e\p\theta(a-r)+\e\w\theta(r-a),
\ee
with the water permittivity $\ew$ and the polymer permittivity $\e\p$. Moreover, the salt screening parameter is defined as $\kappa^2=4\pi\ell_{\rm B}\left(n_{+{\rm b}}+n_{-{\rm b}}\right)$, where $n_{\pm {\rm b}}$ stands for the bulk concentration of the monovalent salt ions. As discussed at the beginning of Sec.~\ref{th}, due to the grand-canonical nature of our formalism, the ion concentrations satisfy automatically the bulk electroneutrality condition, i.e. $n_{+{\rm b}}-n_{-{\rm b}}+q\ce n_{\rm cb}=0$.

\begin{figure*}
\includegraphics[width=1.\linewidth]{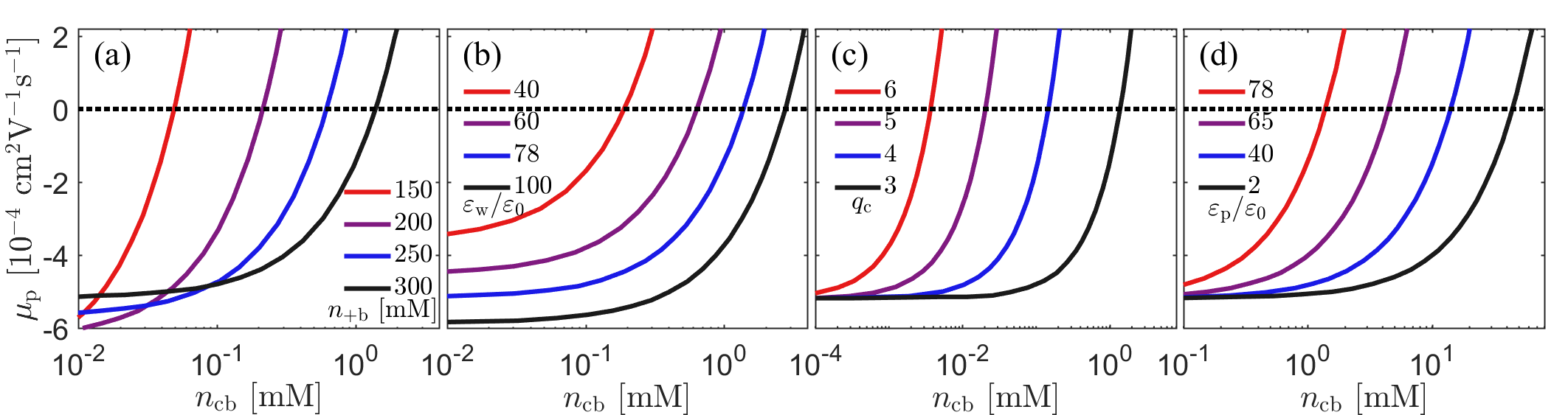}
\caption{(Color online) (a) Polymer mobility $\mu\p$ in Eq.~(\ref{eq15}) versus the multivalent counterion concentration $n_{\rm cb}$ for various values of the (a) monovalent salt concentration $n_{+{\rm b}}$, (b) the electrolyte permittivity $\e\w$, (c) the multivalent charge valency $q\ce$, and (d) the polymer permittivity $\e_{\rm p}$. The model parameters are $a=1.2$ nm, $\delta=0.1$~nm, and $\sigma_{\rm p}=0.783$ $e/{\rm nm}^2$. The charge valency in (a)-(b) and (d) is $q\ce=3$, and the salt concentration in (b)-(d) is $n_{+{\rm b}}=300$ mM.}
\label{fig1}
\end{figure*}

In Eq.~(\ref{eq3}), we introduced the partition function of the multivalent ions
\be\label{eq6}
k\ce(\br)=\theta(r-a)e^{-q\ce\phi_{\rm s}(\br)-q\ce^2\delta G(\br)/2}
\ee
related to the counterion density as $\rho\ce(r)=n_{\rm cb} k\ce(r)$, with the ionic self-energy
\be
\label{eq7}
\delta G(\br)=\lim_{\br'\to\br}\left[G(\br,\br')-G_{\rm b}(\br-\br')\right],
\ee
where $G_{\rm b}(\br-\br')=\ell_{\rm B}/|\br-\br'|$ stands for the electrostatic Green's function in the bulk region. The self-energy~(\ref{eq7}) takes into account two separate electrostatic effects. The first one is the repulsive {\it ionic solvation} force on the multivalent charges originating from the screening deficiency in the salt-free polymer volume. The second effect is the strongly repulsive image-charge forces associated with the dielectric contrast between the polymer and the solvent.

\begin{figure}
\includegraphics[width=1.\linewidth]{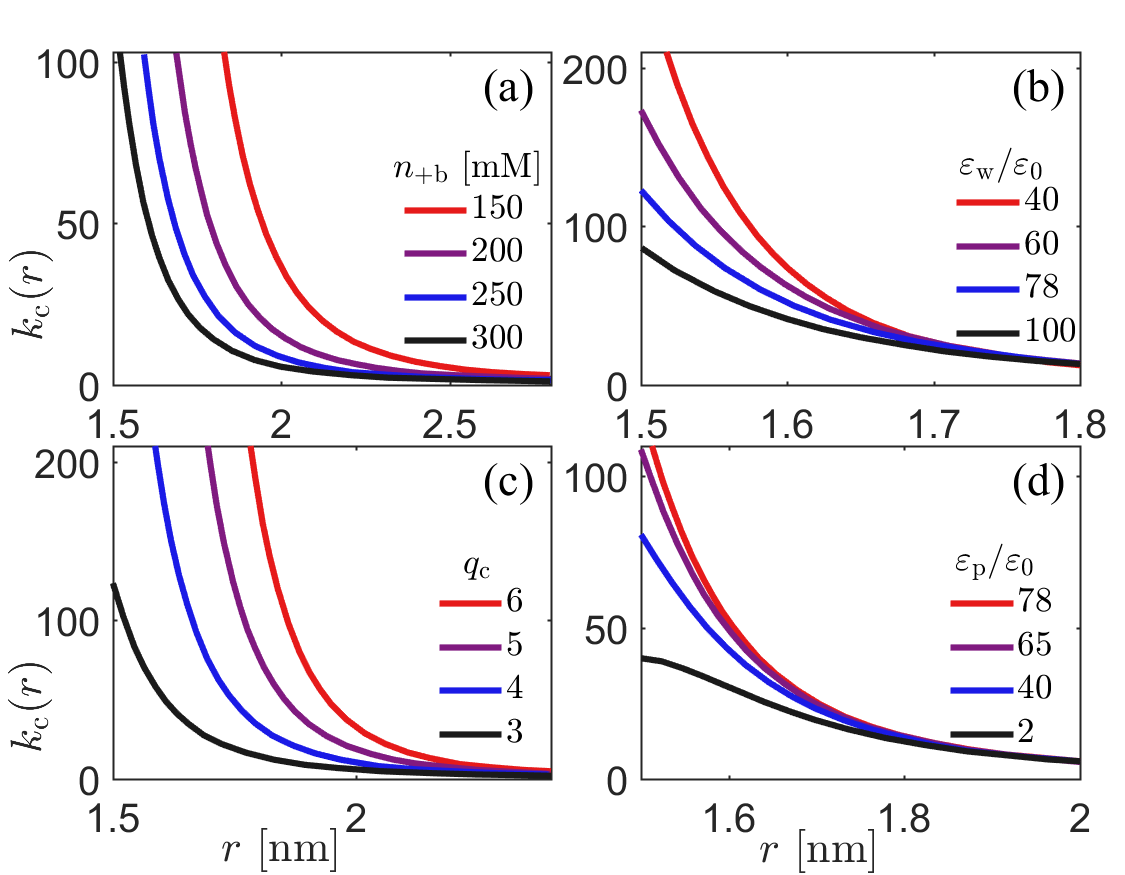}
\caption{(Color online) (a) Dimensionless multivalent ion density $k\ce(r)=n\ce(r)/n_{\rm cb}$ against the separation distance from the polymer axis at the model parameters in Figs.~\ref{fig1}(a)-(d).}
\label{fig2}
\end{figure}

The evaluation of the average potential components~(\ref{eq2})-(\ref{eq3}) requires the knowledge of the Green's function satisfying Eq.~(\ref{eq4}). The latter can be solved by exploiting the cylindrical symmetry of the system. The details of this solution in Fourier space can be found in Refs.~\citenum{Buyukdagli2011,Buyukdagli2021}. The Fourier expansion of the Green's function and its bulk limit read
\bea
\label{eq8}
G(\br,\br')&=&\frac{\ell_{\rm B}}{\pi}\sum_{n=-\infty}^\infty\int_{-\infty}^{\infty}\mathrm{d}k\;e^{in(\phi-\phi')}e^{ik(z-z')}\\
&&\hspace{1.6cm}\times\left[I_n(pr_<)+\Delta_nK_n(pr_<)\right]K_n(pr_>); \nonumber\\
\label{eq9}
G_{\rm b}(\br,\br')&=&\frac{\ell_{\rm B}}{\pi}\sum_{n=-\infty}^\infty\int_{-\infty}^{\infty}\mathrm{d}k\;e^{in(\phi-\phi')}e^{ik(z-z')}\\
&&\hspace{2cm}\times I_n(pr_<)K_n(pr_>),\nonumber
\eea
where we used the modified Bessel functions of the first kind $I_n(x)$ and of the second kind $K_n(x)$, the auxiliary screening parameter $p=\sqrt{\kappa^2+k^2}$, the shortcut notations $r_<={\rm min}(r,r')$ and $r_>={\rm max}(r,r')$ for the radial coordinates, and the dielectric jump function
\be
\label{eq10}
\Delta_n=\frac{pI_n(ka)I'_n(pa)-\gamma kI_n(pa)I'_n(ka)}{-pI_n(ka)K'_n(pa)+\gamma kK_n(pa)I'_n(ka)}
\ee
including the dielectric coefficient $\gamma=\e\p/\ew$. Injecting Eqs.~(\ref{eq8})-(\ref{eq9}) into Eq.~(\ref{eq7}), the self-energy in Eq.~(\ref{eq6}) follows as
\be
\label{eq11}
\delta G(r)=\frac{\ell_{\rm B}}{\pi}\sum_{n=-\infty}^\infty\int_{-\infty}^{\infty}\mathrm{d}k\;\Delta_nK^2_n(pr).
\ee

Substituting now the Green's function~(\ref{eq8}) into Eq.~(\ref{eq2}), the mean-field (MF) potential follows as the standard solution of the linear PB equations around a charged cylinder,
\be\label{eq12}
\phi\s(r)=-\frac{2}{q\ce\kappa\mu_{\rm GC}}\frac{K_0(\kappa r)}{K_1(\kappa a)},
\ee
as expected. Moreover, the correlation component~(\ref{eq3}) of opposite sign responsible for CI becomes
\be
\label{eq13}
\phi\ce(r)=4\pi\ell_{\rm B}q\ce n_{\rm cb}J(r),
\ee
where we introduced the integral function
\bea
\label{eq14}
J(r)&=&K_0(\kappa r)\int_a^r\mathrm{d}r'r' k\ce(r')\\
&&\hspace{1.6cm}\times\left\{I_0(\kappa r')+\frac{I_1(\kappa a)}{K_1(\kappa a)}K_0(\kappa r')\right\}\nonumber\\
&&+\left\{I_0(\kappa r)+\frac{I_1(\kappa a)}{K_1(\kappa a)}K_0(\kappa r)\right\}\nonumber\\
&&\hspace{3mm}\times\int_r^\infty\mathrm{d}r'r' k\ce(r')K_0(\kappa r').\nonumber
\eea
Plugging Eqs.~(\ref{eq12})-(\ref{eq13}) into Eqs.~(\ref{eq1})-(\ref{eq3}), the polymer mobility in Eq.~(\ref{eq0}) finally takes the form
\be
\label{eq15}
\mu\p=-\frac{e\sigma\p}{\kappa\eta}\frac{K_0(\kappa a^*)}{K_1(\kappa a)}+\frac{eq\ce n_{\rm cb}}{\kappa^2\eta}\left[\kappa^2J(a^*)-1\right].
\ee

Equation (\ref{eq15}) is the main result of the present work. We emphasize that to our knowledge this identity is the first analytical beyond-MF mobility formula accounting for the SC correlations induced by the multivalent charges. The first term of Eq.~(\ref{eq15}) corresponds to the MF-level EP mobility, and the second term of opposite sign is the multivalent counterion contribution responsible for mobility reversal. From Eq.~(\ref{eq15}), the critical multivalent cation concentration at the mobility reversal follows as
\be\label{eq16}
n^*_{\rm cb}=\frac{\kappa\sigma\s}{q\ce}\frac{K_0(\kappa a^*)}{K_1(\kappa a)}\frac{1}{\kappa^2J(a^*)-1}.
\ee
In the simplest case, where the no-slip length vanishes, i.e. $a^*=a$, Eqs.~(\ref{eq15})-(\ref{eq16}) simplify to
\bea
\label{eq17}
\mu\p&=&-\frac{e\sigma\p}{\kappa\eta}\frac{K_0(\kappa a)}{K_1(\kappa a)}\\
&&+\frac{eq\ce n_{\rm cb}}{\kappa^2\eta}\left\{\frac{\kappa}{aK_1(\kappa a)}\int_a^\infty\mathrm{d}r'r' k\ce(r')K_0(\kappa r')-1\right\}; \nonumber\\
\label{eq18}
n_{\rm cb}^*&=&\frac{\kappa\sigma\s}{q\ce}\frac{K_0(\kappa a)}{K_1(\kappa a)}\\
&&\times\left\{\frac{\kappa}{aK_1(\kappa a)}\int_a^\infty\mathrm{d}r'r' k\ce(r')K_0(\kappa r')-1\right\}^{-1}.
\eea

\begin{figure*}
\includegraphics[width=1.\linewidth]{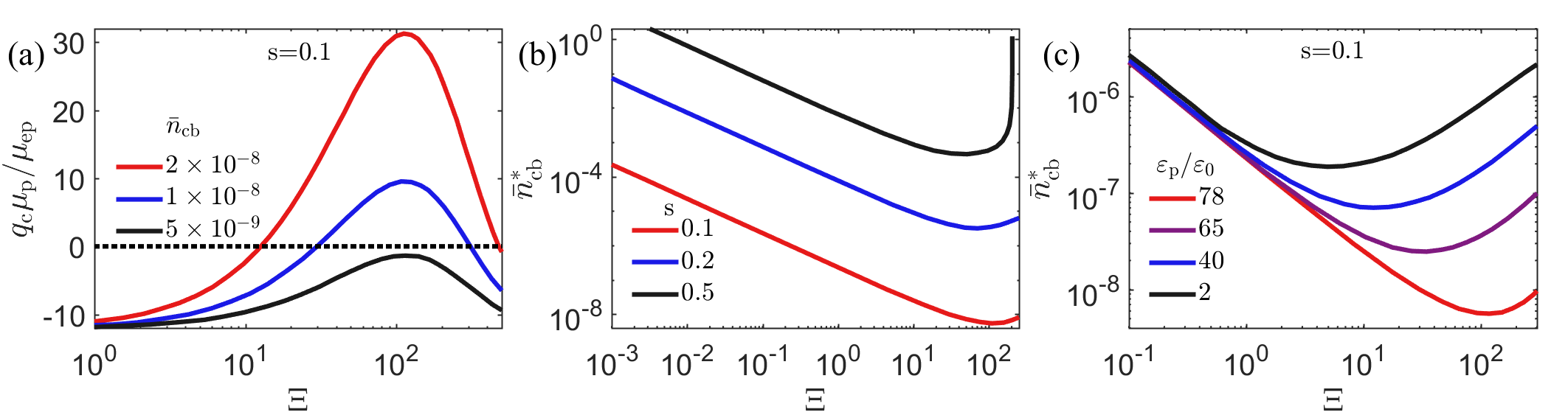}
\caption{(Color online) (a) Polymer mobility~(\ref{eq23}) and (b)-(c) critical concentration~(\ref{eq25}) versus the electrostatic coupling parameter $\Xi$. The polymer radius is $\bba=10.0$ and the no-slip length is $\bar{\delta}=1.0$. The remaining parameters are given in the legends.}
\label{fig3}
\end{figure*}
\subsection{Characterization of the EP mobility inversion}

Figure~\ref{fig1} displays the polymer mobility in Eq.~(\ref{eq15}) versus the multivalent counterion concentration, and Fig.~\ref{fig2} illustrates the multivalent ion density profile in Eq.~(\ref{eq6}) at various model parameters. These mobility and counterion density plots should be interpreted together.

The inspection of Figs.~\ref{fig1}(a)-(b) indicates that the theoretical mobility formula~(\ref{eq15}) agrees qualitatively with the MD simulation results in Figs.~\ref{fig:mobiCurve}(b)-(c). Namely, Figs.~\ref{fig2}(a) and (b) respectively show that the increment of the monovalent salt or the solvent permittivity suppressing the interfacial potential attenuates the counterion adsorption, i.e. $n_{+{\rm b}}\uparrow k\ce(r)\downarrow$ and $\ew\uparrow k\ce(r)\downarrow$. In Figs.~\ref{fig1}(a)-(b), one sees that due to the resulting reduction of charge correlations, the larger the monovalent salt concentration or the liquid permittivity, the larger the multivalent counterion concentration required for the mobility reversal, i.e.  $n_{+{\rm b}}\uparrow n_{\rm cb}^*\uparrow$ and $\ew\uparrow n_{\rm cb}^*\uparrow$.

Figures~\ref{fig1}(c) and~\ref{fig2}(c) indicate that charge valency brings an opposite effect to polymer mobility. Indeed, one sees that multivalent cations of larger valency exhibit a stronger adsorption and higher interfacial density, i.e. $q\ce\uparrow k\ce(r)\uparrow$. Due to the resulting intensification of the surface charge correlations, the higher the charge valency, the lower the counterion concentration at the mobility reversal, i.e. $q\ce\uparrow n_{\rm cb}^*\downarrow$. We note that this trend is equally in agreement with the computational result of Fig.~\ref{fig:mobiCurve}(a).

Finally, we investigate the surface polarization effects associated with the low dielectric permittivity of the polymer on its mobility. Figure~\ref{fig2}(d) shows that for $\e_{\rm p}<\ew$, the repulsive image-charge forces embodied by the self-energy term $\delta G(r)$ of Eq.~(\ref{eq6}) exclude the multivalent counterions from the surface of the polymer, i.e. $\e_{\rm p}\downarrow k\ce(r)\downarrow$. As a result, Fig. \ref{fig1}(d) indicates that the lower the polymer permittivity, the higher the required bulk counterion concentration for the mobility reversal, i.e. $\e_{\rm p}\downarrow n^*_{\rm cb}\uparrow$. We finally note that the present theory does not capture the re-entrance regime of the simulation results in Figs.~\ref{fig:mobiCurve}(a)-(b). This limitation of the SC formalism is a consequence of the virial treatment of the multivalent charges, restricting the validity of the theory to dilute multivalent counterion concentrations.

\subsection{Scaling of the mobility and the reversal concentration}
\label{sc}

In our MD simulation results presented in Sec.~\ref{MD}, the effect of the experimentally controllable physical parameters on the mobility inversion was explained qualitatively in terms of the electrostatic coupling parameter $\Xi$. In order to provide a solid mathematical background supporting this analysis, we investigate here the dependence of the analytically derived EP mobility formula~(\ref{eq15}) on this coupling parameter. To this aim, we first scale all lengths by the GC length as $\bbr=r/\mu_{\rm GC}$, and define the dimensionless salt screening parameter $s=\kappa\mu_{\rm GC}$ and the electrostatic coupling parameter $\Xi=q\ce^2\ell_{\rm B}/\mu_{\rm GC}$. The counterion partition function~(\ref{eq6}) becomes
\be\label{eq19}
k\ce(\bbr)=\theta(\bbr-\bba)e^{-\psi(\bbr)-\Xi U\s(\bbr)},
\ee
with the scaled average potential and self-energy
\bea\label{eq20}
\psi(\bbr)&=&-\frac{2}{s}\frac{K_0(s\bbr)}{K_1(s\bba)};\\
\label{eq21}
U\s(\bbr)&=&\sum_{n=-\infty}^\infty\int_{-\infty}^{\infty}\frac{\mathrm{d}\bak}{2\pi}\;\Delta_nK^2_n(\bap\bbr),
\eea
where we introduced the additional dimensionless parameters $\bba=a/\mu_{\rm GC}$, $\bak=k\mu_{\rm GC}$, $\bap=\sqrt{s^2+\bak^2}$, and
\be
\label{eq22}
\Delta_n=\frac{\bap I_n(\bak\bba)I'_n(\bap\bba)-\gamma\bak I_n(\bap\bba)I'_n(\bak\bba)}{-\bap I_n(\bak\bba)K'_n(\bap\bba)+\gamma\bak K_n(\bap\bba)I'_n(\bak\bba)}.
\ee
In terms of these scaled parameters, the polymer mobility~(\ref{eq15}) takes the form
\be
\label{eq23}
\mu\p=\frac{\mu_{\rm ep}}{q\ce}\left\{-\frac{2}{s}\frac{K_0(s\bba^*)}{K_1(s\bba)}+4\pi\left[I(a^*)-s^{-2}\right]\Xi\;\bn\right\},
\ee
with the dimensionless counterion concentration $\bn=\mu_{\rm GC}^3n_{\rm cb}$, and the auxiliary integral
\bea
\label{eq24}
I(\bbr)&=&K_0(s\bbr)\int_{\bba}^{\bbr}\mathrm{d}\bbr'\bbr' k\ce(\bbr')\\
&&\hspace{1.6cm}\times\left\{I_0(s\bbr')+\frac{I_1(s\bba)}{K_1(s\bba)}K_0(s\bbr')\right\}\nonumber\\
&&+\left\{I_0(s\bbr)+\frac{I_1(s\bba)}{K_1(s\bba)}K_0(s\bbr)\right\}\nonumber\\
&&\hspace{3mm}\times\int_{\bbr}^\infty\mathrm{d}\bbr'\bbr' k\ce(\bbr')K_0(s\bbr').\nonumber
\eea

We first consider the case $\e_{\rm p}=\ew$, where ionic image-charge forces are absent. In Eq.~(\ref{eq23}), one sees that the electrostatic coupling parameter $\Xi$ is involved in the positive correlation component corresponding to the second term in the bracket at two different levels. First, the coupling parameter $\Xi$ comes into play as a linear prefactor scaling the multivalent cation concentration $\bn$. The corresponding dependence on the parameter $\Xi$ originates from the correlations between the polymer charges and the multivalent counterions. These correlations driving the counterion adsorption onto the polymer surface give rise to the polymer CI and the mobility reversal. Indeed, one notes that in the limit of vanishing coupling parameter ($\Xi\to0$), the correlation component in Eq.~(\ref{eq23}) disappears, and the EP mobility tends to its MF counterpart corresponding to the first term in the bracket.

Secondly, Eq.~(\ref{eq24}) shows that $\Xi$ equally contributes to the mobility~(\ref{eq23}) non-linearly as the self-energy magnitude of the ionic partition function in Eq.~(\ref{eq19}). This non-linear contribution takes into account the correlations of the multivalent counterions with their monovalent salt cloud and the polarization charges. These correlations generate the repulsive solvation and image-charge forces suppressing the multivalent counterion adsorption.

Figure~\ref{fig3}(a) shows that this competition gives rise to the non-uniform trend of the polymer mobility~(\ref{eq23}) with respect to the electrostatic coupling parameter. Namely, one sees that at large enough multivalent cation concentrations (blue and red), the negative mobility initially governed by MF electrophoresis rises linearly with $\Xi$, switches from negative to positive at the CI point, and reaches a peak located at $\Xi \approx 100$. In the regime of larger coupling strengths, where the repulsive ionic solvation energy in Eq.~(\ref{eq19}) significantly excludes multivalent counterions from the polymer surface, the resulting weakening of the polymer CI suppresses the reversed mobility and turns the latter from positive back to negative.

From Eq.~(\ref{eq23}), the critical concentration for the mobility reversal follows as
\be\label{eq25}
\bn^*=\frac{1}{2\pi s\Xi}\frac{K_0(s\bba^*)}{K_1(s\bba)}\frac{1}{I(\bba^*)-s^{-2}}.
\ee
Figure~\ref{fig3}(b) shows that due to the amplification of the interfacial counterion adsorption responsible for CI, the rise of $\Xi$ leads to the linear drop of the characteristic counterion concentration~(\ref{eq25}) up to $\Xi=\Xi\ce\approx 100$. At larger $\Xi$ values, where the multivalent counterion adsorption is significantly weakened by the repulsive ionic solvation forces, the characteristic concentration $\bn^*$ reverses its slope and rises quickly with the coupling parameter. Figure~\ref{fig3}(c) indicates that if one also takes into account the low dielectric permittivity of the polymer, the addition of the strongly repulsive image-charge interactions to the ionic solvation forces drops substantially the boundary of the self-energy-dominated coupling parameter regime, i.e. $\e_{\rm p}\downarrow\Xi\ce\downarrow$.

\section{Summary and Conclusions}

In this work, we have combined correlation-corrected transport theory and particle-based numerical simulations to characterize the electrostatic mechanisms behind various experimentally observed features of DNA mobility inversion. Our MD simulations were run in the $NVT$ ensemble. On the other hand, our polymer transport theory was derived in the dressed-ion limit~\cite{Kanduc2011} of the SCPB formalism~\cite{Buyukdagli2020,Buyukdagli2021} incorporating mutually the weak and SC interactions induced by the mono- and multivalent ions, respectively. As this theory is based on the ${\rm \mu}VT$ ensemble, the underlying grand-canonical picture includes the bulk charge reservoir present in the physical system. The main result of this theoretical approach is the identity of Eq.~(\ref{eq15}), providing the first-known-to-us analytical EP mobility formula taking into account the contribution from SC correlations responsible for mobility inversion. We summarize below the main predictions of this analytical formula, equally supported by our numerical simulations, providing qualitatively equivalent results.

The EP mobility formula and its scaled form in Eq.~(\ref{eq23}) indicate -- as anticipated -- that EP polymer mobility reversal is caused by the excess of multivalent counterions adsorbed onto the DNA surface. The strength of this effect originating from the strong coupling of the multivalent counterions and the DNA charges is set by the magnitude of the coupling parameter $\Xi$ in the second term of Eq.~(\ref{eq23}). Our MD simulation results show that as the adsorption excess is accompanied with the CI of the macromolecule, the resulting like-charge coion attraction by the anionic polyelectrolyte comes into play as an additional macroscopic signature of mobility reversal (see Fig.~\ref{fig:dens}). 

We have also investigated the effect of the experimentally controllable system components on the polymer mobility. We found that the increment of the monovalent salt component weakens charge correlations and the multivalent charge adsorption onto DNA, resulting in the suppression of the EP mobility reversal. Consequently, the amount of multivalent counterions $n_{\rm cb}^*$ required for the occurrence of mobility inversion increases with the monovalent salt concentration $n_{+{\rm b}}$, i.e. $n_{+{\rm b}}\uparrow n_{\rm cb}^*\uparrow$. However, as the multivalent counterion-DNA coupling is enhanced by the increase in the ion valency or the decrease of the dielectric permittivity, i.e. $q\ce\uparrow\Xi\uparrow$ and $\ew\downarrow\Xi\uparrow$, the increase of the counterion charge or the decrease of the solvent permittivity lowers the minimum amount of multivalent charges required for mobility inversion, i.e. $q\ce\uparrow n_{\rm cb}^*\downarrow$ and $\ew\downarrow n_{\rm cb}^*\downarrow$. It is noteworthy that the aforementioned predictions are in qualitative agreement with the observations of the EP polymer transport experiments~\cite{Wang2016,Wang2018}.

In the presence of curved interfaces separating water and solvent-free dielectric cavities, such as the surface of DNA with permittivity $\e_{\rm p}\approx 2-5$ in contact with the high permittivity electrolyte, the computational cost associated with the presence of an infinite number of image-charge interactions does not allow the explicit inclusion of the surface polarization effects into the MD simulation framework. However, due to the underlying continuum field representation, our SC-dressed transport theory naturally incorporates the image-charge effects on the polymer mobility. Based on this generality of the theory, we showed that the strongly repulsive polarization forces originating from the dielectric contrast at the polymer-solvent interface repel the multivalent counterions and reduce their coupling to the DNA charges. Consequently, the dielectric cavity created by the low permittivity polymer rises the minimum amount of trivalent counterions required for mobility inversion by almost two orders of magnitude. 

Finally, we have considered two additional peculiarities exclusively accessible by our simulation approach. In Fig.~\ref{fig:mobiCurve}, we showed that the multivalent counterions triggering mobility inversion at dilute concentrations suppress the inversion at large concentrations. The corresponding re-entrance phenomenon previously observed in experiments on DNA condensation~\cite{Hsiao2008,Buyukdagli2017,Truzzolillo2018} calls for verification by EP transport experiments. Additionally, in order to characterize the effect of the typically extended charge structure of multivalent cations, such as spermidine, we investigated the impact of the counterion density on the EP mobility. Our simulations revealed that as the extended structure of the multivalent cations reduces their surface charge density, also resulting in their rotational penalty close to the DNA surface, the finite counterion size weakens the DNA-counterion coupling, therefore suppressing the mobility reversal. Consequently, the characteristic multivalent counterion concentration for mobility reversal increases with the ion charge magnitude. It should be noted that the modeling here simplifies the molecular level dependencies in nucleic acid -- ion interactions, see, {\it e.g.}, \cite{Lipfert2014,Antila2014,Antila2015b}.

In the present study, the MD framework chosen for our numerical computations required the simulation of the charged system in the $NVT$ ensemble, characterized by the constraint of conserved particle number. Future Monte-Carlo simulations using the particle insertion method would enable the simulation of the model in the grand-canonical ensemble, thereby allowing direct quantitative comparison of the theoretical and simulation results. Moreover, our theoretical and computational models neglect the conformational polymer fluctuations. It has been experimentally shown that the stiff polymer approximation holds for polymer lengths extending up to the persistence length of $l_{\rm p}\approx 30-55$ nm~\cite{Brunet2015}. Therefore, the consideration of long polymers necessitates the incorporation of the polymer conformations into the model. The latter is a highly challenging task, and is beyond the scope of the present article.

\section*{SUPPLEMENTARY MATERIAL}
The Supplementary Material consist of assessment of finite-size effects in the MD simulations and additional details regarding convergence of the PPPM electrostatics calculation method in the simulations.

\section*{Acknowledgements}
This work was supported by the Academy of Finland through its Centres of Excellence Programme (2022-2029, LIBER) under project no. 346111 (M.S.). The work was also supported by Technology Industries of Finland Centennial Foundation TT2020 grant (T.A-N. and X.Y.).
 We are grateful for the support by FinnCERES Materials Bioeconomy Ecosystem. Computational resources by CSC IT Centre for Finland and RAMI -- RawMatters Finland Infrastructure are also gratefully acknowledged.

\section*{Author Declarations}
\subsection*{Conflict of Interest}
The authors have no conflicts of interest to disclose.

\section*{Data Availability}
Link to plotted simulations data is provided at https://research.aalto.fi/. If using the open data, we request acknowledging the authors by a citation to the original source (this publication).

\clearpage

\section*{Supplementary Material}

\subsection*{Finite-size effects in molecular dynamics simulations}

In this section, we explore the influence of the size of the simulation box on the ion density and the average potential profiles. A major constraint for the convergence of the system size $V$ is the minimum amount of counterions required to screen the polymer charges. Therefore, according to the identity $N\ce=\rho\ce V$,
low charge concentrations $\rho\ce$ require a large system size to keep the counterion number $N\ce$ above some treshold value needed to compensate for the DNA charges.

The aforementioned effect can be illustrated by comparing the ion number density at the boundary of the simulation box with its bulk value. According to this criterion, finite-size effects can be considered to be negligible if the normalized multivalent ion density defined as $k\ce(r)=n_{\rm c}(r)/n_{\rm cb}$ tends to unity at the system boundary located at $r^*$. As an example, we observed that at high concentrations located in the regime $n_{\rm cb}\gtrsim100$ mM, a simulation box of size $L_x=24$ nm was able to provide a sufficient amount of counterions to compensate the PE charges. Consequently, at this box size, the ion density at the system boundary was sufficiently close to unity ($k\ce(r^*)\gtrsim0.98$) for finite-size effects to be negligible.

\begin{figure}[th]
\begin{centering}
\includegraphics[width=0.98\linewidth]{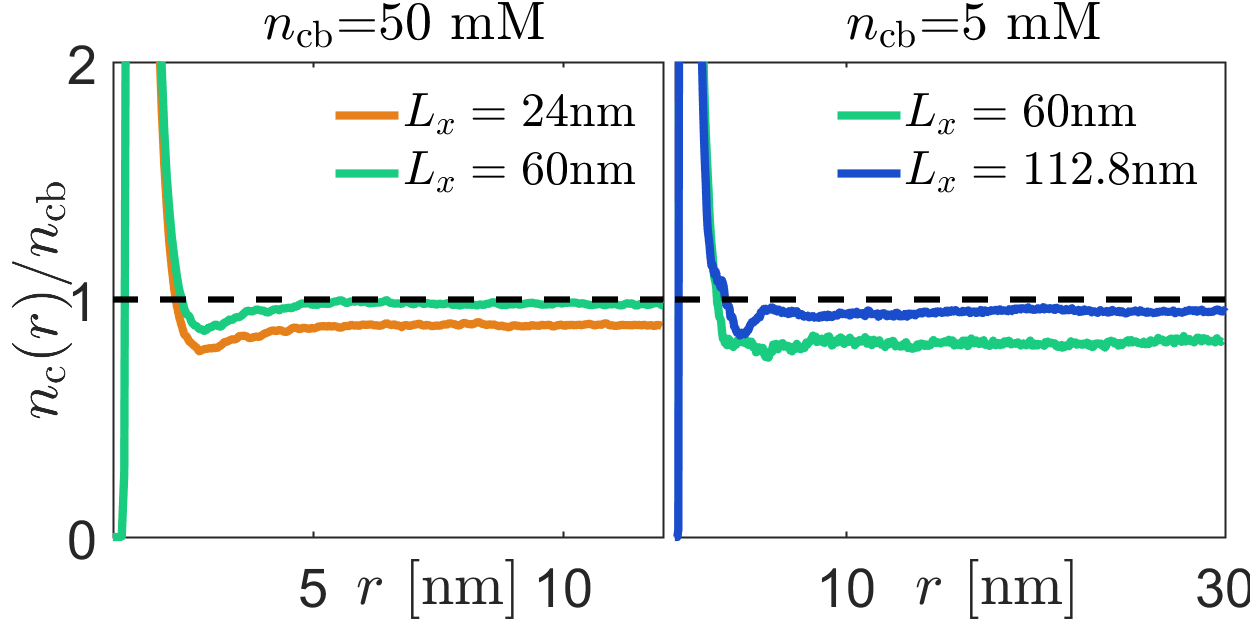}
\par\end{centering}
\caption{Normalized trivalent ion density at the bulk concentrations $n_{\rm cb}=50$~mM (left) and $n_{\rm cb}=5$~mM (right), and for different dimensions of the simulation box (see the legends) displaying reduction in the finite-size effects at $L_x\gtrsim60$ nm (left) and $L_x\gtrsim112.8$ nm (right). The relative permittivity and the longitudinal system size are $\varepsilon_{\rm w}/ \varepsilon_0=78$ and $L_z=20$ nm. 
\label{fig:dens_FS}}
\end{figure}

\begin{figure}[th]
\begin{centering}
\includegraphics[width=1\linewidth]{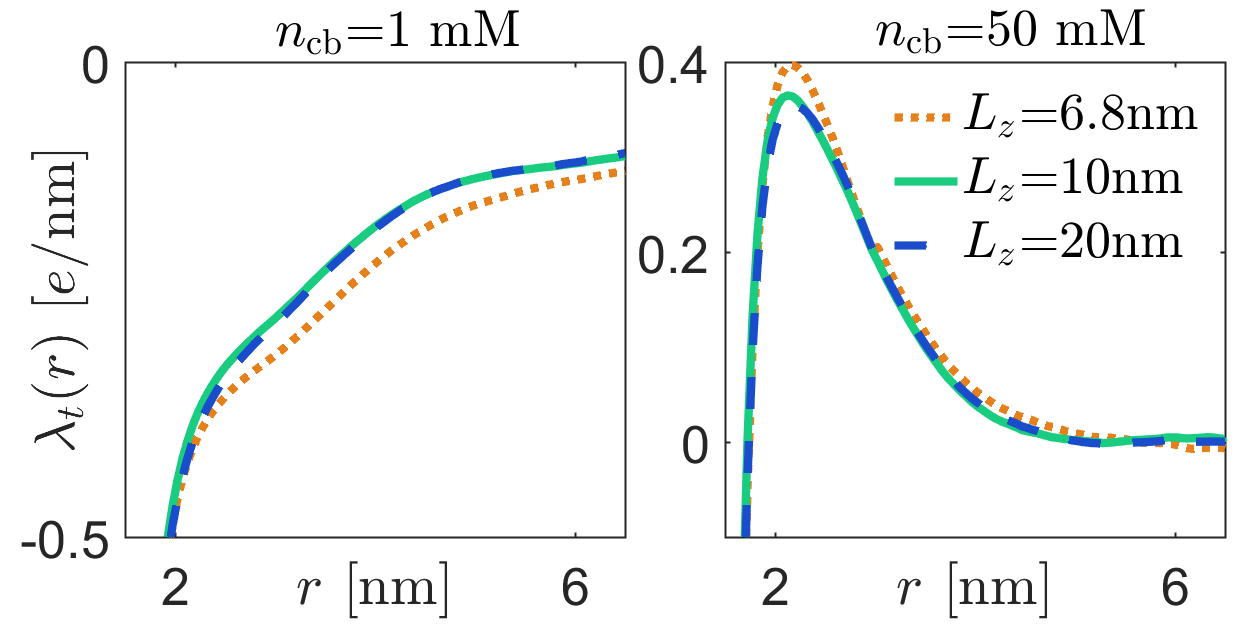}
\par\end{centering}
\caption{Cumulative charge density with trivalent counterions of bulk concentrations $n_{\rm cb}=1$~mM (left) and $n_{\rm cb}=50$~mM (right), and different longitudinal system sizes (see the legend) indicating the numerical convergence at $L_z\gtrsim10$ nm. The relative permittivity is $\varepsilon_{\rm w}/ \varepsilon_{0}=78$, and the transverse system size is $L_x=112.8$ nm (left) and $L_x=60$ nm (right).
\label{fig:dens_FS_Lz}}
\end{figure}

The left plot of Fig.~\ref{fig:dens_FS} shows that for the same box size $L_x=24$ nm but at the lower counterion concentration $n_{\rm cb}=50$ mM (orange curve), due to the decrease of the counterion number below the critical amount required to screen the DNA charges, the normalized ion density at the system boundary drops below unity.
One sees that by increasing the box size to $L_x=60$ nm (green curve), the density at the boundary increases to unity, indicating the elimination of the visible finite-size effects from the simulations. 
In the right panel of Fig.~\ref{fig:dens_FS}, one sees that decreasing further the counterion concentration down to $n_{\rm cb}= 5$ mM, the box size $L_x=60$ nm becomes insufficient to satisfy the thermodynamic limit in the transverse direction (green curve). In this case, the increase of the system size up to $L_x=112.8$ nm allowed again to restore the thermodynamic limit by rising the boundary value of the normalized ion density significantly close to unity (blue curve). 
In Table~\ref{tab:box_size}, we report the characteristic system sizes enabling the substantial suppression of the finite-size effects by keeping the boundary value of the normalized ion density above the value of $0.95$.

An additional finite-size effect is the finite boundary value of the electrostatic potential. In the physical system, the bulk electroneutrality condition leads to a vanishing potential in the bulk reservoir. However, in the simulation box of finite transverse dimensions, the electrostatic potential always has a finite value at the boundaries of the system. In order to minimize this finite-size effect, the transverse size of the box should be chosen significantly larger than the characteristic screening lengths of the potential. Thus, the elimination of this effect required us to choose the system size such that the latter satisfies the condition $L_x\gg\mu_{\rm GC}$ for the salt-free liquid and $L_x\gg\kappa^{-1}$ in the presence of salt. The corresponding values are displayed in Table~\ref{tab:box_size}.

\begin{table}[ht]
\begin{centering}
\begin{tabular}{||c |c |c |c |c ||}
\hline
$n_{\rm cb}$~[mM] & $L_x$~[nm] & max $\kappa^{-1}$~[nm] & $\mu_{\rm GC}$~[nm] & $d^{\rm s}$ [nm] \tabularnewline
\hline
$0.1 - 0.5$ & 240 & 25 & 0.1 & 3.0 \tabularnewline
\hline
$1 - 5$ & 112.8 & 7.9 & 0.1 & 3.0 \tabularnewline
\hline
$10 - 50$ & 60 & 2.5 & 0.1 & 1.5 \tabularnewline
\hline
$>100$ & 24 & 0.8 & 0.1 & 1.5 \tabularnewline
\hline

\end{tabular}
\par\end{centering}
\caption{Characteristic lengths and parameters: the horizontal box size $L_x$, the screening length $\kappa^{-1}$, the GC length $\mu_{\rm GC}$, and the PPPM distance $d^{\rm s}$ for various trivalent salt concentrations and the relative permittivity $\ew/ \varepsilon_0=78$.
\label{tab:box_size}}
\end{table}

Finally, in order to eliminate finite-size effects associated with the length of the polymer, we had to identify the characteristic polymer length above which the results remain unaffected by an increase in the molecule length. In Fig.~\ref{fig:dens_FS_Lz}, we show that the increase of the polymer length above $L_z\approx10$ nm does not lead to a significant change of the cumulative charge density. Thus, in order to ensure that our ion condensation results are free of finite length effects, we always set the polymer length to $L_z=20$ nm.

\subsection*{Additional details regarding convergence of the PPPM electrostatics calculation method}
\label{pppm}

We provide here details regarding convergence considerations of the particle–particle particle-mesh (PPPM) method used to evaluate the pairwise electrostatic interactions between the charges in the polymer-ion complex. 
The PPPM approach splits the electrostatic Coulomb interactions into a short-range and a long-range component. The short-range component defined as the interactions between each charge and its neighbors located within a distance $d^{\rm s}$ is evaluated directly in the method.
The long-range interactions of the central charge with the other charges located at a larger distance $r>d^{\rm s}$ are evaluated by mapping the interacting particles into a grid and carrying-out a fast Fourier transform~\cite{Brown2012}. The Key parameter for the convergence of the calculation is $d^{\rm s}$. Grid spacing is generated automatically from $d^{\rm s}$ by LAMMPS. %and Fourier space cut-off. 
For computational efficiency, the parameter $d^{\rm s}$ is often in coarse-grained modelling  set to a value slightly larger than the particle size. 
However, for elevated local charge density such as in this study, the choice can lead to loss of convergence when $d^{\rm s}> 4$~nm. We carefully checked the ion distribution convergence in our work. 
Based on the outcome of the convergence checking, in our simulations, we use $d^{\rm s}=1.5$ nm for system sizes in the range $L_x \leq 60$~nm. 
In larger systems where the simulations are significantly more time consuming if extensive weight is on the mesh part, we use a larger value of $d^{\rm s}=3$ nm. This allows fast numerical computation by keeping practically the same accuracy. The ion distribution remains the same as $d^{\rm s}=1.5$ nm.
The corresponding values of the parameter $d^{\rm s}$ are reported in Table~\ref{tab:box_size}.
The parameter of relative error was set to $10^{-5}$ and the stencil size was set to 5. 
No other PPPM calculation parameters were modified from LAMMPS algorithm implementation defaults.

\bibstyle{approve.bib}
\bibliography{ref_mobiRev.bib}

\end{document}